\documentclass[journal=nalefd,manuscript=letter,layout=twocolumn]{achemso}

\usepackage[T1]{fontenc}
\usepackage{graphicx}
\usepackage{dcolumn}
\usepackage{siunitx}
\DeclareSIUnit{\barpressure}{bar}
\DeclareSIUnit{\langmuir}{L}
\DeclareSIUnit\angstrom{\protect \text {Å}}
\usepackage{amsmath}
\usepackage{xcolor}
\DeclareGraphicsExtensions{.pdf,.png}

\AbstractOn
\SectionsOn

\author{A. De Vita}
\altaffiliation{These authors contributed equally.}%
\affiliation{Dipartimento di Fisica, Università degli Studi di Milano, Via Celoria 16, I-20133 Milano, Italy\looseness=-1}%
\alsoaffiliation{Istituto Officina dei Materiali (IOM)-CNR, Laboratorio TASC, in Area Science Park, S.S.14, km 163.5, I-34149 Trieste, Italy\looseness=-1}%
\author{R. Sant}%
\altaffiliation{These authors contributed equally.}%
\affiliation{ESRF, The European Synchrotron, 71 Avenue des Martyrs, CS40220, 38043 Grenoble Cedex 9, France\looseness=-1}%
\author{V. Polewczyk}
\affiliation{Istituto Officina dei Materiali (IOM)-CNR, Laboratorio TASC, in Area Science Park, S.S.14, km 163.5, I-34149 Trieste, Italy\looseness=-1}%
\author{G. van der Laan}
\affiliation{ Diamond Light Source, Harwell Science and Innovation Campus, Didcot, Oxfordshire
OX11 0DE, UK\looseness=-1}%
\author{N. B. Brookes}
\affiliation{ESRF, The European Synchrotron, 71 Avenue des Martyrs, CS40220, 38043 Grenoble Cedex 9, France\looseness=-1}%
\author{T. Kong}
\affiliation{Department of Chemistry, Princeton University, Princeton, New Jersey 08540, United States\looseness=-1}%
\author{R. J. Cava}
\affiliation{Department of Chemistry, Princeton University, Princeton, New Jersey 08540, United States\looseness=-1}%
\author{G. Rossi}
\affiliation{Dipartimento di Fisica, Università degli Studi di Milano, Via Celoria 16, I-20133 Milano, Italy\looseness=-1}%
\alsoaffiliation{Istituto Officina dei Materiali (IOM)-CNR, Laboratorio TASC, in Area Science Park, S.S.14, km 163.5, I-34149 Trieste, Italy\looseness=-1}%
\author{G. Vinai}
\email{vinai@iom.cnr.it}%
\affiliation{Istituto Officina dei Materiali (IOM)-CNR, Laboratorio TASC, in Area Science Park, S.S.14, km 163.5, I-34149 Trieste, Italy\looseness=-1}%
\author{G. Panaccione}
\affiliation{Istituto Officina dei Materiali (IOM)-CNR, Laboratorio TASC, in Area Science Park, S.S.14, km 163.5, I-34149 Trieste, Italy\looseness=-1}%

\title{Evidence of temperature-dependent interplay between spin and orbital moment in van der Waals ferromagnet VI\textsubscript{3}}


\begin{document}

\begin{abstract}

Van der Waals materials provide a versatile toolbox for the emergence of new quantum phenomena and the fabrication of functional heterostructures. Among them, the trihalide VI\textsubscript{3} stands out for its unique magnetic and structural landscape. Here we investigate the spin and orbital magnetic degrees of freedom in the layered ferromagnet VI\textsubscript{3} by means of temperature-dependent x-ray absorption spectroscopy and x-ray magnetic circular and linear dichroism. We detect localized electronic states and reduced magnetic dimensionality, due to electronic correlations. We furthermore provide experimental evidence of (a) an unquenched orbital magnetic moment (up to $\num{0.66\pm0.07}$) in the ferromagnetic state, and (b) an instability of the orbital moment in proximity of the spin reorientation transition. Our results support a coherent picture where electronic correlations give rise to a strong magnetic anisotropy and a large orbital moment, and establish VI\textsubscript{3} as a prime candidate for the study of orbital quantum effects.

\end{abstract}

\section{Main text}

The progress in fundamental research on next-generation magnetic materials is currently spearheaded by two-dimensional (2D) van der Waals (vdW) materials. These compounds have been garnering a large share of attention, as they exhibit a wide array of layer-dependent magnetic ordering 
\cite{Butler2013,Samarth2017,Mounet2018,Burch2018,Gibertini2019}; tailoring vdW materials may pave the way for applications in the ever-growing field of spintronics \cite{Ankit2021}.

Among them, trihalide VI\textsubscript{3} has been extensively studied in the last years \cite{Son2019,Tian2019,Kong2019,Yang2020,Zhang2022,Bergner2022,Wang2023}. At variance with the Ising ferromagnet CrI\textsubscript{3}, VI\textsubscript{3} is a canted layered ferromagnet with strong magnetocrystalline anisotropy, displaying in its bulk form a structural transition from monoclinic to rhombohedral below $T_s=\SI{79}{\K}$ and two ferromagnetic ones, one at the Curie temperature $T_{\mathrm{FM1}}=\SI{50}{\K}$ and the other at $T_{\mathrm{FM2}}\approx\SI{35}{\K}$ with a variation of spin canting direction and stacking order of layers. The proximity of structural and FM transitions in a narrow temperature interval hints at a relevant interplay between crystal structure and magnetic order. Interestingly, once VI\textsubscript{3} is reduced to few monolayers, $T_{\mathrm{FM1}}$  depends on the number of layers \cite{Lin2021}; moreover, further clues suggest that VI\textsubscript{3} hosts strong magnetoelastic interactions \cite{Dolezal2019}. The importance of orbital phenomena in solids \cite{Khomskii2021} motivates this line of research on VI\textsubscript{3} also due to its reduced dimensionality. However, to date the understanding of this physics is elusive and mostly limited to magnetometric and structural characterizations. 

For this reason, a temperature-dependent spectroscopic investigation of the spin and orbital degrees of freedom is of paramount importance. X-ray magnetic circular dichroism (XMCD) and x-ray natural linear dichroism (XNLD) are ideal techniques for this research. Structural changes influencing the transition metal centres can be monitored by studying the orbital magnetic moment, highly dependent on the local symmetry of the crystal; additionally, the spin magnetic moment reveals the details of the long-range magnetic order, as well as identifying the universality class of the material and its spin dimensionality.

In this Letter, we firstly report the variation of the magnetic and orbital signals in VI\textsubscript{3} crystals, below and above the two FM transitions, by combining XMCD and XNLD measurements. We reveal the presence of a large circular dichroic signal at low temperature, with a clear discontinuity near $T_{\mathrm{FM2}}$ and disappearance above $T_{\mathrm{FM1}}$, following a tricritical behavior and compatible with a reduced dimensionality. Linear dichroic measurements disclose a mixed ground state, consistent with a partially unquenched orbital moment supported by cluster calculations, strongly quenched in the proximity of $T_{\mathrm{FM2}}$. Our results confirm the presence of a high orbital moment in VI\textsubscript{3} at low temperature and reveal an orbital instability near $T_{\mathrm{FM2}}$, corroborating the hypothesis of a strong coupling between orbital and magnetic degrees of freedom. These findings underline the relevance of orbital magnetization in VI\textsubscript{3} among vdW crystals and reinforce the importance of quantum spin-orbit entanglement.

\begin{figure*}
\includegraphics[width=0.9\linewidth]{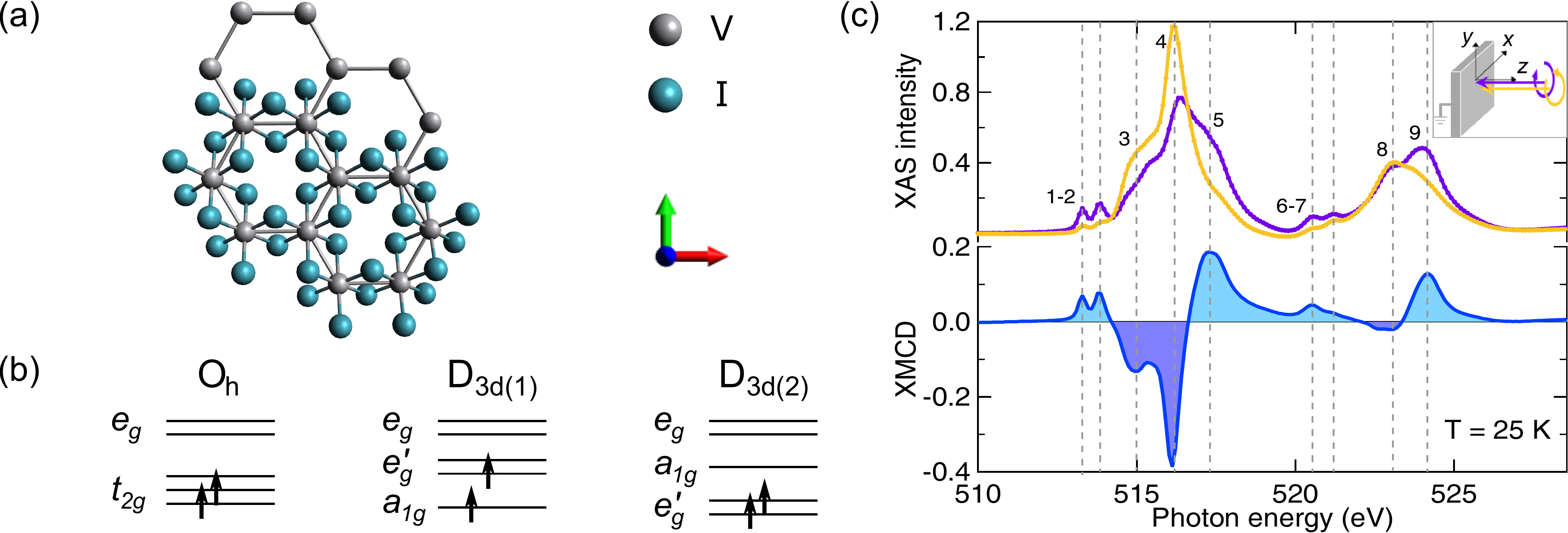}
\caption{\label{fig:Fig1}(a) Top pictorial view of a VI\textsubscript{3} layer. Each V centre is coordinated by six I atoms to form VI\textsubscript{6} octahedra. (b) Crystal field splitting and electron filling for O\textsubscript{h} and D\textsubscript{3d} symmetry; in the latter case 3\textit{d} levels are split by elongation (1) or compression (2) along the trigonal axis, resulting in opposite ordering of $a_{1g}$ and $e'_{g}$ orbitals. (c) \textit{Top:} XAS spectra across the V \textit{L}\textsubscript{2,3} edges, acquired with left and right circularly polarized photons at \SI{25}{\K}. The inset shows the experimental geometry. \textit{Bottom:} Corresponding XMCD.}
\end{figure*}

The experiment has been performed at the XMCD endstation of the ID32 beamline at the European Synchrotron Radiation Facility (ESRF) \cite{Kummer2016,Brookes2018}. Commercially available VI\textsubscript{3} crystals stored in Ar atmosphere ($<0.5$ ppm O\textsubscript{2}, $<0.5$ ppm H\textsubscript{2}O) have been transferred in inert static atmosphere and cleaved under N\textsubscript{2} flow inside the loadlock chamber, to expose the (0001) crystallographic plane of the clean surface without contaminating the highly hygroscopic surface; the sample is held in ultra-high vacuum (UHV, pressure $<\SI{3e-10}{\milli\barpressure}$). The structure and 3\textit{d} electronic configuration of VI\textsubscript{3} are represented in Fig. \ref{fig:Fig1}a-b. X-ray absorption spectroscopy (XAS) spectra have been acquired in total electron yield and normalized by the intensity collected on a Au mesh in front of the sample stage. The beam has almost \SI{100}{\percent} degree of linear and circular polarization, and the setup has resolving power better than 5000. XMCD spectra are the difference between left- and right- circularly polarized light spectra, measured at remanence after zero-field cooling from room temperature to \SI{25}{\K} following application of \SI{0.5}{\tesla} out-of-plane magnetic field, large enough to saturate the magnetization \cite{Son2019}; the sample temperature spans the range from \SIrange{25}{65}{\K}. The beam spot diameter at sample position at normal incidence is \qtyproduct{100x100}{\um}. XNLD spectra are the difference between vertically and horizontally polarized light spectra, measured at \SI{20}{\K}, at grazing incidence ($\theta=\ang{70}$) and at remanence. All spectra are normalized by the average of the two light polarizations.

Multiplet cluster calculations have been performed using the \textit{Quanty} code \cite{Haverkort2016} to simulate XNLD signal at \SI{20}{\K}. The trigonal crystal field perturbation has been set to zero, modeling instead the trigonal distortion via hybridization potentials \smash{$V_{a_{1g}}$} and \smash{$V_{e'_g}$}. The simulation parameters and the role of trigonal distortion/hybridization in the orbital ordering of VI\textsubscript{3} are discussed in detail elsewhere \cite{Sant2023}.



XAS and XMCD spectra across the V \textit{L}\textsubscript{2,3} edges at $T=\SI{25}{\K}$ are presented in Fig. \ref{fig:Fig1}c. The XAS lineshape shows a complex multiplet structure in agreement with that reported for similarly measured VI\textsubscript{3} spectra, confirming the 3+ valence state \cite{DeVita2022,Hovancik2023}. The \textit{L}\textsubscript{3} edge is dominated by five spectral features: specifically at \SI{513.3}{\eV} and \SI{513.8}{\eV} (1-2, in the pre-edge region), \SI{515}{\eV} (3), \SI{516.2}{\eV} (4) and \SI{517.3}{\eV} (5). The \textit{L}\textsubscript{2} peak displays two primary features at \SI{523.1}{\eV} (8) and \SI{524.1}{\eV} (9); between the two edges, the \textit{L}\textsubscript{2} pre-edge features at \SI{520.5}{\eV} and \SI{521.2}{\eV} (6-7) are the broader counterparts of the (1-2) \textit{L}\textsubscript{3} pre-edges.

\begin{figure*}
\includegraphics[width=16cm]{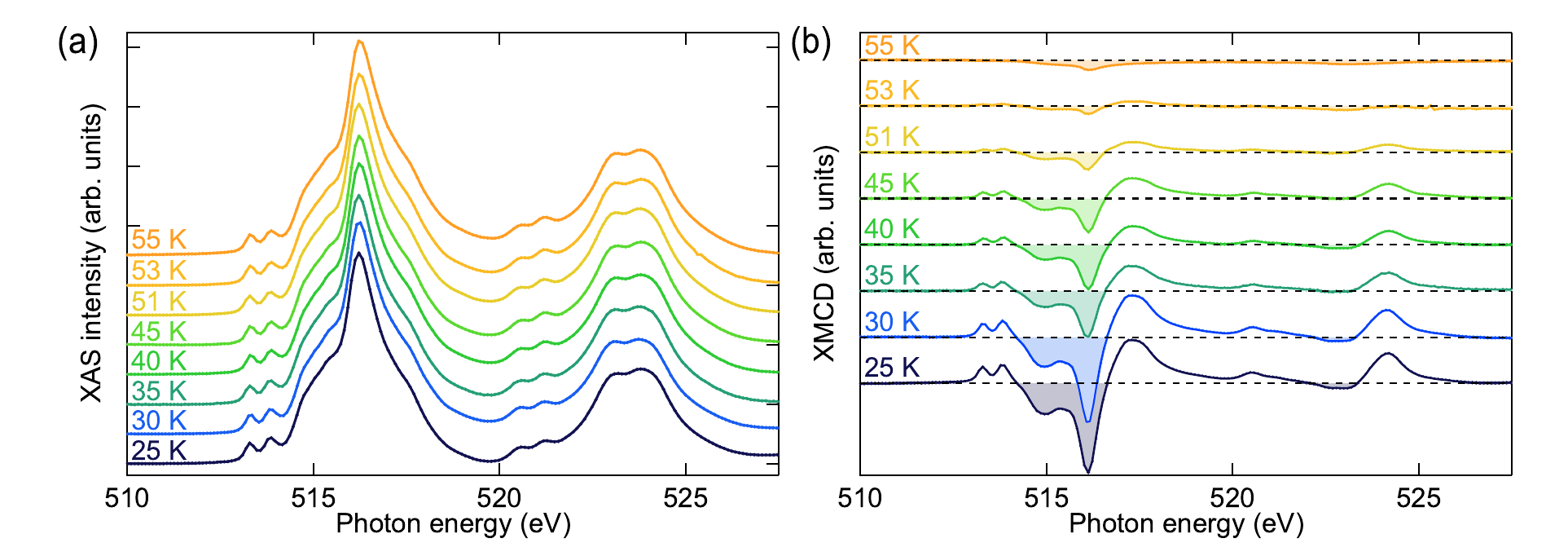}
\caption{\label{fig:Fig2}(a) XAS average spectra across the V \textit{L}\textsubscript{2,3} edges as a function of temperature. (b) Corresponding XMCD.}
\end{figure*}

Concerning the XMCD spectrum, the features of \textit{L}\textsubscript{3} and \textit{L}\textsubscript{2} edges present both negative (3, 4, 8) and positive (5, 9) dichroism, owing to the combined weak spin-orbit coupling (SOC) and strong hybridization typical in V compounds \cite{DeGroot1990,Abbate1993,Matsuura2015,Zhang2019,Vinai2020,Schmitz2020,Maganas2020}. On the other hand, the pre-edge features 1-2 and 6-7, which are associated with transitions from V 2\textit{p} to the unoccupied V 3\textit{d} $t_{2g}^3$ state at the V\textsuperscript{3+} site, present narrow structures, hinting at a localized nature of 3\textit{d} states in the valence band. In the case of \textit{L}\textsubscript{3} pre-edges, the XMCD intensity almost corresponds to the intensity of the right circular light, signature of a large asymmetry in the spin population. In addition, both edges exhibit an uncommonly large positive dichroic signal. The positive sign for the \textit{L}\textsubscript{3} pre-edge, where no shoulder affects the sign of the dichroic signal, is attributed to the final state, \textit{i.e.} an $\uparrow$ electron in the previously partially filled $t_{2g\uparrow}$ states (see Fig. \ref{fig:Fig1}b), to which a corresponding negative one would be expected for the \textit{L}\textsubscript{2} pre-edge \cite{Matsuura2015}. We attribute this lack in inversion to the large degree of hybridization of V\textsuperscript{3+} in the crystal: indeed, the long tail of the \textit{L}\textsubscript{3} main peak at $\sim\SI{517}{\eV}$ (5) extends up to the \textit{L}\textsubscript{2} pre-edge region, shadowing the sign reversal of the pre-edge peak. 


The large dichroic signal on the \textit{L}\textsubscript{3} main peak ($\approx\num{0.4}$) is a clear indication of FM ordering. In order to study the dependence of the magnetic character on temperature, we performed the same measurements while raising the temperature up to above $T_{\mathrm{FM1}}$, which are plotted in Fig. \ref{fig:Fig2}a-b. The temperature increase produces no qualitative modification in the features of both XAS and XMCD spectra. The XMCD lineshape reported in Fig. \ref{fig:Fig2}b shows that magnetic order is preserved up to \SI{51}{\K}, whereas from \SI{53}{\K} onward the signal intensity becomes comparable with the sensitivity limit of the beamline, losing information on the details of the overall spectrum. This indicates that the Curie temperature lies between \SI{51}{\K} and \SI{53}{\K}, in agreement with literature \cite{Tian2019,Son2019,Kong2019,Gati2019}.

\begin{figure}
\includegraphics[width=0.9\linewidth]{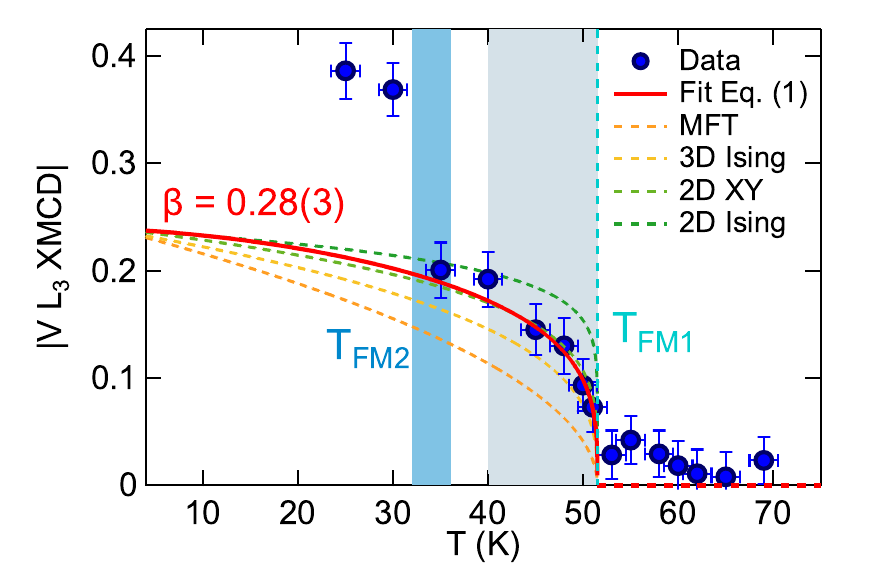}
\caption{\label{fig:Fig3}XMCD maximal intensity at the V \textit{L}\textsubscript{3} edge as a function of temperature, and visual comparison with theoretical models. The solid red line plots the fit following Eq. \ref{eq:cw}, in the temperature range highlighted by the gray area; dotted lines trace curves for mean field ($\beta=0.5$), 3D Ising ($\beta=0.33$), 2D XY ($\beta=0.23$) and 2D Ising ($\beta=0.125$) models. The teal dotted line marks $T_{\mathrm{FM1}}$ as retrieved from the fit, the light blue shaded area indicates $T_{\mathrm{FM2}}$.}
\end{figure}

The XMCD signal allows an element-sensitive quantitative evaluation on the magnetic spin and orbital moments via sum rules analysis \cite{Chen1995}. However, calculations of the spin moment contribution become inadequate for early transition metals, where the large overlap between \textit{L}\textsubscript{3} and \textit{L}\textsubscript{2} edges prevents a clear-cut distinction \cite{Obrien1994}. Therefore, we follow the evolution of VI\textsubscript{3} magnetic character by plotting the XMCD intensity at the \textit{L}\textsubscript{3} main peak, as displayed in Fig. \ref{fig:Fig3}. A first observation is the abrupt reduction of the signal between \SI{30}{\K} and \SI{35}{\K}. This discontinuity may be explained in terms of the nature of the FM transition. In previous studies \cite{Gati2019} a weak discontinuity of the heat capacity appears at $T_{\mathrm{FM2}}$: additionally, since a second-order transition takes place at $T_{\mathrm{FM1}}$, thermodynamic considerations prevent the one at $T_{\mathrm{FM2}}$ to display a second-order character \cite{Yip1991}. We thus infer that a weak first-order transition at this temperature would result in a discontinuity of the order parameter. Moreover, a structural transition from triclinic to monoclinic has been reported at $T_{\mathrm{FM2}}$, together with an in-plane rotation of the canted V spins \cite{Dolezal2019,Koriki2021,Hao2021}; this could justify the variation of dichroic signal below $T_{\mathrm{FM2}}$, as the concurrent change of crystal structure and magnetic easy axis could modify the projection of the magnetic moment along the out-of-plane direction.

Around $T_{\mathrm{FM1}}$, the trend of the XMCD signal can be assessed in order to extract information on the spin dimensionality. To do so, we fitted the curve in Fig. \ref{fig:Fig3} with the Curie-Weiss law 

\begin{equation}\label{eq:cw}
    m \sim (\tau)^\beta,
\end{equation}

where $\tau = (T_{\mathrm{FM1}}-T)/T_{\mathrm{FM1}}$, and $\beta$ and $T_{\mathrm{FM1}}$ are free parameters of the fit. The procedure has been performed for $\tau \leq 0.1$, \textit{i.e.} in a region close to the transition, to ensure the validity of the formula. The fit yields values of $T_{\mathrm{FM1}} = \SI{51.5\pm0.5}{\K}$ and $\beta = \num{0.28\pm0.03}$. The critical exponent $\beta$ governing the order parameter gives indications on the universality class of the material, being $\beta=\num{0.23}$ for dimensionality $n=2$ and $\beta=\num{0.5}$ for dimensionality $n=3$ (\textit{cf.} the dotted curves exemplifying different models in Fig. \ref{fig:Fig3}). Our results suggest that VI\textsubscript{3} magnetic character is denoted by a reduced spin dimensionality, as expected for a quasi-2D crystal with weak inter-plane interactions \cite{Taroni2008}, in agreement with what found on other vdW materials \cite{Zaiyao2018,Bedoya2021,Wildes2006} and by magnetometric measurements on VI\textsubscript{3} \cite{Liu2020,Lin2021}.

The extracted value of $\beta$ and the first-order FM transition at $T_{\mathrm{FM2}}$ point at the tricritical mean field model as a good description of the critical behavior of VI\textsubscript{3}. In the VI\textsubscript{3} $p-T$ phase diagram \cite{Gati2019}, the two FM transitions merge at a triple point ($p_{c1}$, $T_{c1}$) with increasing pressure: our measurements further support the hypothesis that this is also a tricritical point.


In addition to XMCD, we performed XNLD on V \textit{L}\textsubscript{2,3} edges to retrieve information on the orbital ordering. We showed that the pre-edge features previously identified (Fig. \ref{fig:Fig1} and discussion) result from transitions to unoccupied $t_{2g}$ states; in particular, the trigonal distortion splits the $t_{2g}$ level into $a_{1g}$ and $e'_{g}$, with mostly out-of-plane and in-plane character respectively. In an XNLD experiment with grazing angle geometry, horizontally (vertically) polarized photons interact mostly with out-of-plane (in-plane) orbitals: it is thus possible to discern which level is occupied from the change in the spectral weight of the pre-edge features. 

\begin{figure}
\includegraphics[width=0.9\linewidth]{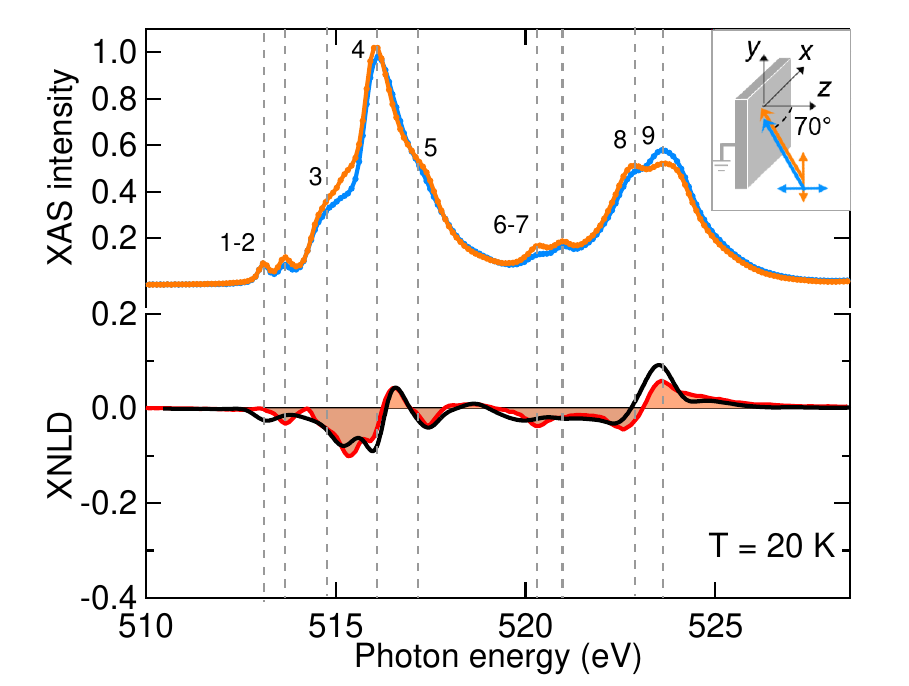}
\caption{\label{fig:Fig4}\textit{Top:} XAS spectra across the V \textit{L}\textsubscript{2,3} edges, acquired with linearly polarized photons at \SI{20}{\K}. The inset shows the experimental geometry. \textit{Bottom:} calculated (solid black line) and measured (red shaded area) XNLD.}
\end{figure}

XAS spectra acquired with linearly polarized light, and the corresponding XNLD, are displayed in Fig. \ref{fig:Fig4}. We reveal a negative dichroism in both pre-edge regions (1-2 and 6-7): the signal derived from the excitation induced by linear light polarization vector $E \perp c$ (vertical polarization) is enhanced compared to $E \parallel c$ (horizontal polarization), consistently with the XNLD reported by our group \cite{Sant2023}. By means of cluster calculations, the XNLD at \SI{20}{\K} is simulated in the framework of the ligand-field multiplet theory. In Fig. \ref{fig:Fig4}, the resulting curve (solid black line) obtained with hybridization parameters \smash{$V_{a_{1g}}=\SI{1.097}{\eV}$} and \smash{$V_{e'_g}=\SI{1.103}{\eV}$} is in excellent agreement with experimental data, indicating an \smash{$a_{1g}^1 e_{g}^{'1}$} ground state. Furthermore, as a comparison with a similar V\textsuperscript{3+} system, in V\textsubscript{2}O\textsubscript{3} the pre-edge intensity increases when the $a_{1g}$ orbital is unoccupied and $E \parallel c$ \cite{Park2000}. Our observation suggests that in the initial state the $a_{1g}$ orbital is filled: this is also in agreement with a recent ARPES characterization, where the $a_{1g}$ orbital lies at the top of the valence band and displays strong photoemission intensity \cite{DeVita2022}, and with the large orbital moment measured by Hovancik et al. \cite{Hovancik2023}. We note however that the resulting linear dichroism is quite small compared to e.g. the one calculated for V\textsubscript{2}O\textsubscript{3}, indicating that the energy difference between \smash{$e_{g}^{'2}$} and \smash{$a_{1g}^1 e_{g}^{'1}$} ground states is small, and tiny lattice distortions and/or partial charge transfer could stabilize a mixed electronic configuration \cite{Sant2023}.

\begin{figure}
\includegraphics[width=0.9\linewidth]{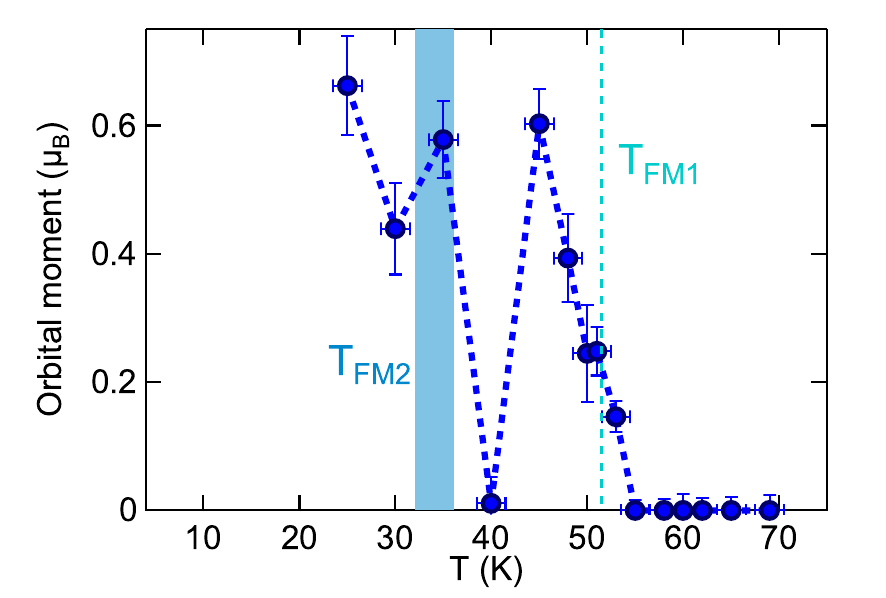}
\caption{\label{fig:Fig5}Orbital magnetic moment on V\textsuperscript{3+} as a function of temperature. The green dotted line and the teal shaded area indicate the two FM transition temperatures.}
\end{figure}

The stabilization of an \smash{$a_{1g}^1 e_{g}^{'1}$} ground state drives the appearance of a high orbital moment in FM VI\textsubscript{3}. To assess the effect of temperature, we calculated the orbital moment from XMCD data displayed in Fig. \ref{fig:Fig2} by applying the orbital sum rule \cite{Chen1995}, as shown in Fig. \ref{fig:Fig5}. The value at \SI{25}{\K} ($L_z=\num{0.66\pm0.07}$) is comparable with the other reported result \cite{Hovancik2023} for a $\SI{50}{\percent}$ occupancy of each possible ground state (\smash{$e_{g}^{'2}$} and \smash{$a_{1g}^1 e_{g}^{'1}$}). However, as the temperature increases we notice a strong quenching of the orbital moment at $\SI{40}{\K}$, which then recovers and gradually goes to zero close to $T_{\mathrm{FM1}}$. If we consider that $T_{\mathrm{FM2}}$ lies between \SI{32}{\K} and \SI{36}{\K} \cite{Gati2019,Dolezal2019,Marchandier2021,Soler2022}, the proximity of the sharp disruption of the orbital moment to $T_{\mathrm{FM2}}$ indicates a connection to the critical dynamics of the transition: the orbital instability is indeed a signature of a change in the anisotropy constant of the lattice \cite{Garcia2000}, and has been demonstrated to be a precursor of spin reorientation transitions \cite{Removic2003}. Changes in the magnetic anisotropy accompanied by a spin reorientation have also been reported in other vdW materials exhibiting a similar magnetic anisotropy, such as Fe\textsubscript{4}GeTe\textsubscript{2} \cite{Seo2020,Wang2023}.

The presence of an unquenched orbital moment has deep implications in the physics of low-dimensional materials. 3\textit{d} transition metals display quenched orbital moment and weak SOC (in the order of few tens of millielectronvolts); however, strong SOC may appear when electron correlations lead to a spin-entangled state, as suggested in the Ising metallic system Fe\textsubscript{1/4}TaS\textsubscript{2} \cite{Ko2011}, or recently in the 2D vdW antiferromagnet FePS\textsubscript{3} \cite{Lee2023}, where large orbital moment and SOC ultimately give rise to a sizeable magnetic anisotropy. VI\textsubscript{3}, a system with a large unquenched moment where SOC cannot be neglected without hampering the interpretation of experimental spectra \cite{DeVita2022}, has been proposed as a candidate material where electronic correlations enhance the orbital magnetization \cite{Zhou2021}. Therefore, future efforts to understand the microscopic origin of the large magnetic anisotropy in VI\textsubscript{3} should focus on the orbital degree of freedom, in view of spintronics applications.  


In summary, in this work we present and interpret temperature-dependent XMCD measurements, with the support of XNLD and cluster calculations, to assess the spin and orbital degrees of freedom in VI\textsubscript{3}. XMCD reveals a magnetic state with strong (up to $\approx\num{0.4}$) dichroism, whose lineshape is not qualitatively affected by temperature, and well-defined pre-edge features, signature of localized states. 

The XMCD maximal intensity at the \textit{L}\textsubscript{3} edge shows an abrupt decrease at the first-order FM transition at $T_{\mathrm{FM2}}$, and displays a critical behavior following an exponent $\beta=\num{0.28\pm0.03}$, suggesting a tricritical mean field model and a reduced dimensionality.

The XNLD spectrum exhibits a sizeable negative linear dichroism on the pre-edges, implying an at least partially filled $a_{1g}$ orbital in the ground state and non-zero orbital moment: further analysis from temperature-dependent XMCD spectra confirm a large (up to $L_z=\num{0.66\pm0.07}$) orbital moment, and unveil the presence of a large orbital instability near $T_{\mathrm{FM2}}$. This is interpreted in terms of changes in the magnetic anisotropy of the crystal, disrupting the orbital moment and kickstarting the subsequent spin reorientation. Our study demonstrates that VI\textsubscript{3} hosts exotic quantum properties and will bolster advanced investigations into vdW 2D ferromagnets.

\section{Acknowledgments}
We acknowledge the European Synchrotron Radiation Facility for provision of beamtime on the ID32 beamline. G.P. acknowledges financial support from PNRR MUR project PE0000023-NQSTI. This work has been performed in the framework of the Nanoscience Foundry and Fine Analysis (NFFA-MUR Italy Progetti Internazionali) project (www.trieste.NFFA.eu). 

\section*{Author Contributions}

A.D.V., R.S., G.V. and G.P. designed the research; A.D.V., R.S., V.P., N.B. and G.V. performed the experiment; A.D.V. and G.V. analyzed data; R.S. and G.v.d.L. performed the simulations; R.J.C. and T.K. provided samples for early measurements; G.R. and G.P. provided support with funding and supervision; A.D.V. and G.V. wrote the paper with contributions from all authors. All authors have given approval to the final version of the manuscript.

A.D.V. and R.S. equally contributed to this work.



\bibliography{bibliography}

\end{document}